\documentclass{article}

\usepackage{PRIMEarxiv}
\usepackage{booktabs}
\usepackage{etoolbox}

\providecommand{\keywords}[1]{\par\vskip 0.5em\noindent\textbf{Keywords:} #1}
\AtBeginEnvironment{tabular}{\small}

\usepackage[utf8]{inputenc} % allow utf-8 input
\usepackage[T1]{fontenc}    % use 8-bit T1 fonts
\usepackage{hyperref}       % hyperlinks
\usepackage{url}            % simple URL typesetting
\usepackage{booktabs}       % professional-quality tables
\usepackage{amsfonts}       % blackboard math symbols
\usepackage{nicefrac}       % compact symbols for 1/2, etc.
\usepackage{microtype}      % microtypography
\usepackage{fancyhdr}       % header
\usepackage{graphicx}       % graphics
\graphicspath{{media/}}     % organize your images and other figures under media/ folder
\usepackage{enumitem}
\usepackage[font=small,labelfont=bf]{caption}
\usepackage{minted}

\usepackage{lineno}
\title{Asymmetric Communication: Large Language Models and Language Games
\thanks{An earlier version of this paper is available on SSRN under the
title \textit{Large Language Models and Language Games: Asymmetric Communication} (\url{https://ssrn.com/abstract=6678558}).
The present version revises the title, abstract, and conceptual framing.}
}

\author{
  Enzo Fenoglio\\
  Dept. Computer Science \\
  University College London\\
  66-72 Gower Street, London WC1E 6EA, UK\\
  \texttt{e.fenoglio@ucl.ac.uk} \\
}

\begin{document}
\maketitle
\setcounter{footnote}{0}
\begin{abstract} 
Contemporary AI discourse routinely attributes to language models properties they cannot bear: general intelligence as substrate-independent cognition, hallucination as cognitive failure, agency as autonomous goal-pursuit, sentience as emergent inner life, alignment as goal synchronization. This paper argues that these are instances of a single category mistake — properties constituted within human communicative practice are projected onto the machine side of the interaction — and supplies the structural account that explains why. Human–LLM interaction constitutes a new type of language game, one whose defining property is that a single side carries all normative activity. We call this configuration asymmetric communication since model outputs circulate communicatively, being produced, interpreted, and incorporated into further exchanges, without the system undertaking commitments, bearing entitlements, or performing the assessment on which discursive standing depends. Three conditions define the asymmetry: (i) correctness is enforced exclusively by the receiver; (ii) accountability is borne by human participants alone; and (iii) the practical standing of any output depends entirely on human uptake. These conditions are structural rather than empirical and hold independently of capability — more powerful models raise the stakes of misattribution without altering its structure. The framework draws its apparatus from Wittgenstein (meaning enacted in shared practices), Luhmann (communication completed on the receiver's side), Esposito (algorithmic contingency sufficient for uptake), and Brandom (normative scorekeeping as the source of discursive standing). Applied to these five narratives, it reclassifies each as a receiver-side phenomenon, grounds guardrails as structural necessities rather than manifestations of machine moral agency, and yields a direct implication for AI governance. Because responsibility cannot migrate to systems that occupy no normative position, alignment is constraint engineering within human institutions, not goal synchronization between agents.
\end{abstract}  




% ==================================== SEction 1 ===================
%\pagebreak
\section{Introduction}
Recent advances in large language models (LLMs) have generated intense debate across scientific, industrial, and public domains. Interpretations commonly split between two poles: (i) LLMs as highly capable statistical systems that reproduce linguistic regularities without understanding~\cite{Bender_Gebru_2021} or, more cautiously, as tools whose apparent competence does not justify strong cognitive attributions~\cite{Floridi_2023}, and (ii) LLMs as emerging cognitive agents to which one may ascribe reasoning, intention, or even sentience, sometimes framed in terms of existential risk~\cite{Cosmo_2022,Metz_Schmidt_2023}. The persistence of this split suggests that the central difficulty is not only empirical evaluation, but the conceptual frame used to interpret what LLM behavior can and cannot warrant.

This paper argues that implicit assumptions about intelligence, meaning, and agency structure contemporary AI debates. When these assumptions remain unexamined, fluent generation, adaptive responses, and apparent reasoning are readily over-read as evidence of human-like cognition. The result is a recurrent confusion of categories. Epistemic and intentional properties are attributed to technological performance, producing inflated expectations and misplaced concerns. 

We propose an alternative perspective grounded in a distinction between communicative competence and epistemic agency. LLMs generate linguistically coherent artifacts by modelling regularities in human linguistic practice. Meaning, interpretation, and responsibility remain on the human side of the interaction. Tokens do not carry meaning on their own; meaning arises through human uptake, interpretation, and assessment within ongoing practices. Humans establish the criteria of correctness, determine whether outputs count as reasons or evidence, and bear responsibility for the consequences of their uptake. Human--machine exchanges involving LLMs are therefore intrinsically asymmetric. By \textit{asymmetric communication}\footnote{The term \textit{asymmetric communication} appears in human-robot interaction research to describe a behavioral asymmetry. The robot addresses its interlocutor as a human while the human addresses the robot as a machine \cite{kawakubo_2024}. The present paper uses the same term in a structural and constitutive sense — meaning and normative accountability reside exclusively on the human side as a consequence of what participation in a language game requires, not as a contingent feature of current technology or user perception.} we mean a form of interaction in which LLMs can generate utterances, but the norms that make those utterances count as correct or incorrect, justified or unjustified, responsible or irresponsible remain on the human side. These norms are enforced through the assessment of reasons and commitments (Section~\ref{subsec:communication-asymmetry}).

Our asymmetric communication framework reclassifies several topics often treated as paradoxical. For example, \textit{hallucinations} become cases of fluent generation that lacks epistemic entitlement, rather than failures of reasoning; claims about artificial general intelligence often reflect projections of human cognitive categories onto systems optimized for communicative adequacy rather than autonomous understanding; and agency and alignment concerns shift from imagined machine intention to conditions of human deployment, legitimation, and accountability.

The conceptual approach adopted draws on four complementary philosophical sources. Wittgenstein's notion of language games~\cite{Wittgenstein_1953} emphasizes that meaning and correctness are constituted within shared practices rather than guaranteed by internal representations alone. Luhmann's theory of communication~\cite{Luhmann_1995} foregrounds how communicative events are completed through receiver-side interpretation, decoupling communicative success from assumptions about the sender's inner states. Esposito's account of artificial communication~\cite{esposito_2017} extends this framework to algorithmic systems, showing how they can participate in communicative processes without possessing understanding or intentionality. Brandom's inferentialism~\cite{Brandom_2000} provides the normative infrastructure that underlies the other three for his account of meaning as participation in the game of giving and asking for reasons makes explicit the normative practice of attributing commitments and entitlements that grounds the asymmetry between human and artificial participants.   Together, these perspectives provide a framework for understanding LLM outputs as communicatively usable without attributing to the system human-like intentionality or epistemic standing. The framework does not intervene in the engineering debate over what AI systems should become, nor does it adjudicate between competing architectures or positions on the pace of AI development. Its aim is more fundamental. It identifies the structural conditions under which human--LLM interaction becomes meaningful, responsibility is located, and normative practice is sustained. These conditions are irreducibly located on the human side of the interaction, not as a contingent feature of current technology, but as a consequence of what participation in a language game requires. The practical implication is not a restriction on AI deployment, but a reorientation of how accountability, governance, and institutional design should be understood in socio-technical systems where artificial systems enter communication without thereby becoming bearers of normative standing.

The remainder of the paper proceeds as follows. Section~2 presents related work and current assumptions about intelligence, language, and agency. Section~3 presents the conceptual foundations derived from philosophy of language, communication theory, and inferentialism. Section~4 develops the principle of human--machine asymmetry and its implications for meaning and responsibility. Section~5 reinterprets common AI narratives, including artificial general intelligence, hallucination, agentic AI, emotional projection, and alignment, through this framework. Section~6 concludes with implications for future research and discourse.

% ===================================== Section 2
\section{Related Work}
\label{sec:related_work}
Research on large language models (LLMs) spans multiple conceptual pathways with different assumptions about \textit{intelligence}, \textit{language}, and \textit{agency}. This section maps five dominant trajectories that structure current debates and makes explicit where the paper intervenes conceptually.

\subsection{The Replication Premise}
A dominant tradition treats intelligence as a computational phenomenon that can, in principle, be reproduced through sufficiently expressive architectures and training regimes. Both technical and philosophical programs~\cite{russell_2019, bostrom_2014} and safety-oriented framings~\cite{amodei_2016, christiano_2017} share this premise, differing mainly on how human-level capability should be achieved and at what risk. Within this tradition, influential critiques challenge the LLM-centric route without abandoning the broader replication ambition. LeCun argues that LLMs lack grounding, planning, and physical-world interaction, advocating instead for architectures built around world models and embodied learning~\cite{lecun_2022,Browning_2022}. Karpathy emphasizes that LLMs are fundamentally next-token predictors, and that treating their outputs in terms of \textit{language} and \textit{understanding} misleads conceptual interpretation~\cite{karpathy_2024}. A parallel line of critique emphasizes the epistemic and social risks of treating fluent text generation as evidence of understanding or authority, famously articulated in the \textit{stochastic parrots} argument~\cite{Bender_Gebru_2021}. Despite their disagreement on engineering strategy and normative emphasis, these positions commonly assess progress in relation to human cognitive capacities. Our approach does not contest these technical debates but questions the replication premise as the default interpretive frame.

\subsection{Meaning, Understanding, and the Language Question}
\label{subsec:meaning-understanding}
A substantial philosophical literature asks whether symbol manipulation can constitute understanding, frequently using AI as a test case. Searle's Chinese Room argument establishes the canonical syntax/semantics divide~\cite{searle_1980}, while Harnad's symbol grounding problem extends the challenge to the conditions under which symbols acquire meaning~\cite{harnad_1990}. This line of critique predates the current systems. Dreyfus argued across several decades that symbolic and connectionist approaches alike misconstrue intelligence as rule-following or pattern recognition abstracted from situated being-in-the-world~\cite{Dreyfus_1992}, and Brooks drew the engineering consequence, proposing that intelligent behavior be built without internal representation at all~\cite{Brooks_1991}. The present framework inherits this anti-representationalist lineage but locates what representation cannot supply differently, not in embodied coping — a machine-side remedy that the world-model program has since incorporated (Section~\ref{subsec:world-models}) — but in normative standing, which no architectural provision can confer. LLMs have renewed these debates under new empirical conditions, sharpening the question of whether linguistic competence entails semantic competence or intentionality.

A growing set of works engages Wittgensteinian themes---language games, rule-following, use---in relation to LLMs. Mallory argues that Wittgenstein's anti-foundationalism, his rejection of meaning as an inner or reified entity, and his critique of appeals to understanding as an explanation of linguistic performance together help to clarify how LLMs can exhibit linguistically coherent behavior without thereby possessing human-like understanding~\cite{mallory_2023}.  O'Connor examines whether LLMs can be considered fellow participants in language games, concluding that the question cannot be resolved by behavioral evidence alone~\cite{oconnor_2025}. Coeckelbergh applies Wittgensteinian language game  analysis to technology more broadly, showing that evaluating technological systems requires attending to the normative practices in which they are embedded~\cite{Coeckelbergh_2017}. Mollema compares LLMs to Calvino's literature machines, arguing that while LLMs instantiate language users in a functional sense, ascribing psychological predicates to them remains a functionalist temptation that the Wittgensteinian framework does not support~\cite{mollema_2024}. A recent contribution by Bottazzi et al. diagnoses a \textit{bewitchment} induced by LLM fluency, arguing through Wittgenstein that LLMs disrupt genuine communicative understanding by substituting statistical regularity for the dynamic constancy that human linguistic practice requires~\cite{Bottazzi_2025}. These contributions sharpen the conceptual question but share a common orientation. They ask whether LLM performance licenses not only attributions of understanding or linguistic participation, but in some cases even the claim that genuine communication has occurred at all --- in short, whether the machine is on the right side of the Wittgensteinian line. 

Formal and game-theoretic reinterpretations of language games~\cite{noguerialonso_2024}, contrasts between social-pragmatic and formal-semantic perspectives~\cite{vartziotis_2026}, and analyses of linguistic intentionality in LLMs~\cite{grindrod_2024} pursue the same question from different angles. Other related works also interrogate whether careful language use around LLMs can mitigate attribution errors~\cite{Shanahan_2024}, and Browning and LeCun argue from a linguistic perspective that LLM outputs cannot constitute understanding because they lack grounding in world experience~\cite{Browning_2022}.

Our contribution shifts this center of gravity entirely. The question of whether LLMs understand is, within the framework developed here, secondary to the question of where meaning is constituted in human--LLM interaction. The decisive locus is not an internal property of the generator but the receiver-side conditions under which meaning and agency are attributed in human practice. This reorientation changes what counts as the explanatory target. It explains why Wittgenstein-only approaches, however precise, leave the receiver-side structure under-theorized since they diagnose the machine's limitations without providing a positive account of the communicative structure of human--LLM interaction.

\subsection{Normative Accountability and Moral Agency}
\label{subsec:normative-accountability}
A parallel literature addresses the conditions under which AI systems can bear moral agency, responsibility, or accountability. This tradition is less concerned with whether LLMs understand than with whether they can be held to normative standards and what happens to accountability when artificial systems are deployed in consequential settings.

Brandom's inferentialism provides a useful philosophical articulation of this trajectory, though it is rarely cited explicitly within it~\cite{Brandom_1994}. His account of meaning as participation in the game of giving and asking for reasons establishes that normative statuses---commitments, entitlements, and their assessment through scorekeeping---are irreducibly social achievements, grounded in practices of reciprocal assessment and accountability. This inferentialist account helps explain why artificial systems, on the present view, cannot bear normative standing since they produce outputs that enter human scorekeeping practices as objects of assessment without undertaking scorekeeping themselves.

Within the moral agency literature, two broad positions emerge:
(i) \textit{accommodating views} hold that AI systems can in principle
instantiate the relevant capacities for moral agency --- rationality,
interactivity, and normative judgment --- provided their functional
organization is sufficiently complex~\cite{Floridi_2004};
(ii) \textit{skeptical or eliminativist views} argue that current artificial systems fail to meet the conditions for moral agency because they lack consciousness, moral understanding, intentionality, or the ability to act for reasons rather than merely in accordance with them~\cite{Himma_2009,Veliz_2021}.
A recent treatment grounds this restrictive position in Arendt's account
of responsibility and Strawson's reactive attitudes, arguing that
authorship requires accountability and that LLMs, lacking moral agency,
cannot bear either~\cite{ukpaka_2025}. These positions are genuinely opposed on whether consciousness is necessary for moral agency, yet they share a prior assumption: that the question is whether the system possesses or lacks the properties constitutive of agency. The present framework relocates the question. Whether an artificial system instantiates or fails to instantiate agency-conferring capacities is secondary to whether it occupies a position in a normative practice — which, as Section~\ref{sec:human-machine-communication} argues, is not a capacity a system can possess or lack but a standing conferred within human scorekeeping. 

Shanahan's careful analysis of the language used around LLMs reaches a similar conclusion from a different direction. Namely, the attribution of understanding, belief, and intention to LLMs reflects linguistic habits rather than warranted inferences~\cite{Shanahan_2024}. Overall, these contributions correctly identify that accountability cannot migrate to machines. Their limitation is that they provide no positive account of what happens communicatively when humans interact with LLMs --- they establish what LLMs are not without theorizing what the interaction is. Floridi's \textit{agency without intelligence}~\cite{Floridi_2023} offers a valuable functional characterization of artificial agency, while leaving outside its scope the communication-theoretic structure through which LLM outputs acquire meaning and standing in human practice.

\subsection{Empirical Self-Modeling and Introspection}
\label{subsec:empirical-selfmodeling}
A technically distinct trajectory examines whether LLMs possess functional analogs of self-awareness, introspection, and metacognition. This empirical literature poses the most direct challenge to our framework. If LLMs can genuinely monitor and report on their own internal states, the claim that normativity resides exclusively on the human side requires careful qualification.

Recent research documents a range of functional self-modeling capabilities. Models demonstrate some ability to estimate their own knowledge and predict their own behavior in hypothetical scenarios~\cite{laine_2024, kadavath_2022}. Models finetuned to exhibit specific behavioral propensities can describe those propensities when asked, suggesting privileged access to their own learned dispositions~\cite{betley_2025, plunkett_2025}. Models show partial ability to recognize their own previously generated outputs~\cite{panickssery_2024}, though this result is sensitive to prompt design~\cite{davidson_2024}. Most directly, Lindsey~\cite{Lindsey_2025} investigates introspection by injecting representations of known concepts into model activations and measuring their influence on self-reported states, finding that current models possess a limited, functional form of introspective awareness that is nonetheless highly unreliable and context-dependent.

These findings are significant but do not cross the philosophical line the present framework draws. The capabilities identified in this literature operate at the causal level; they document statistical correlations between internal states and self-reports, and between training signals and behavioral propensities. A system may causally model aspects of its own operation and produce reliable self-reports without thereby entering the normative space of mutual accountability in which such reports count as commitments open to challenge, justification, and revision. They do not establish normative self-awareness, intentional rule-following, or participation in a practice of giving and asking for reasons. As Lindsey notes, models possess a playbook for acting like introspective agents regardless of whether they are~\cite{Lindsey_2025}. The definitional debate within the literature is revealing. Researchers disagree on whether introspection requires grounded causal access to internal states~\cite{comsashanahan_2024}, privileged self-access beyond what third parties could infer~\cite{song_2025}, or merely reliable self-modeling~\cite{binder_2024}. This instability reflects the difficulty of applying concepts grounded in human normative practice to systems that do not inhabit such practices. Functional self-modeling is consistent with a system that generates statistically coherent completions conditioned on its internal state. It does not constitute scorekeeping, commitment, or entitlement in Brandom's sense. What appears as introspection from the outside is a sophisticated behavioral effect that the human receiver interprets as such --- which is precisely what our framework predicts. It is worth making explicit what would, and would not, count as crossing the line in question. The criterion is not behavioral but positional. To undertake a commitment is to stand within a practice of reciprocal assessment in which one's performances can be challenged, in which one can be sanctioned, and in which one bears the consequences of one's entitlements being withdrawn. No behavioral evidence can satisfy a positional criterion, and this is by design rather than by evasion; the framework deliberately declines to treat participation in normative practice as a capability that empirical benchmarks could detect, because the rule-following considerations (Section~\ref{subsec:wittgenstein}) establish that no fact about an isolated system determines normative status. Whether this positional criterion could in principle be satisfied by artificial systems embedded in future institutional arrangements is a substantive question the present paper leaves open; what it rules out is the inference from functional self-modeling, at any capability level, to normative standing.

\subsection{Systems-Theoretic and Sociological Accounts of Artificial Communication}
A distinct pathway treats communication as a social process whose success does not require access to (or agreement about) the sender's inner states. Luhmann formulates communication as a three-part selection process---information, utterance, understanding---completed on the receiver side, decoupling communicative success from psychological assumptions about the producer~\cite{Luhmann_1995}. Building on this, Esposito introduces the notion of \emph{artificial communication}, arguing that algorithms can function within communicative systems without possessing understanding, producing socially meaningful effects by stabilizing expectations under contingency~\cite{esposito_2017, Esposito_WEB_2022}. This is the enabling insight on which the present framework most directly depends. If communicative success is constituted on the receiver's side, the question of whether the producer understands becomes secondary, and an artificial system can participate in communication without thereby being credited with a mind. No account examined in this paper grants as much, and none stands as close to the position developed here.

Its limit lies exactly where its strength does. Luhmann secures receiver-side completion by bracketing normativity as an explanatory category. Communication reproduces itself as a system of selections, and whether any given completion carries entitlement, commitment, or standing is not a question the theory is built to answer; Esposito inherits this bracketing. The result explains how an artificial utterance can enter communication, but not why some completions count as correct, justified, or accountable and others do not --- the specifically normative character of receiver-side uptake. That gap is what a Wittgensteinian account of rule-governed practice and Brandom's inferentialism are introduced to fill (Sections~\ref{subsec:brandom},~\ref{subsec:syntgesis}). Recent work extends the systems-theoretic lens to contemporary AI. Keenan et al.\ reframe explainable AI as a communicative rather than epistemic achievement~\cite{keenan_2024}, and Morton analyses LLMs explicitly as artificial communication partners~\cite{morton_2025}, but both share the same silence on normativity. Empirical work on multi-agent LLM settings sharpens the issue: Ashery et al.~ document conventions emerging among interacting models~\cite{ashery_2025}, a phenomenon easily read as machine-side norm-formation. On the present account, it is not, since such conventions are stabilized patterns whose standing as norms still depends on human uptake at some stage of oversight and evaluation (Section~\ref{subsec:agentic}), not scorekeeping undertaken by the systems themselves. Our framework therefore extends this pathway rather than adopting it, integrating the Luhmann--Esposito account of receiver-side completion with a Wittgensteinian emphasis on use and rule-governed practice.

\subsection{Distinctive Contribution of the Present Framework}
\label{subsec:position}
The five trajectories surveyed above exhibit the common tendency to frame the explanatory question primarily inside the machine. Computationalist approaches ask whether the system replicates cognition. Philosophical treatments of meaning and understanding ask whether the system participates in the relevant linguistic practices. Normative accountability frameworks ask whether the system can bear moral agency or responsibility. Empirical self-modeling research asks whether the system has genuine access to its own internal states. Systems-theoretic accounts ask whether the system functions as a communication partner. In each case, the machine is the primary explanatory target.

Our approach relocates the question. The decisive locus is not an internal property of the generator but the interaction structure --- specifically, the receiver-side conditions under which meaning, agency, and normative standing are attributed in human practice. This reorientation is not a minor adjustment. It changes what counts as an explanatory target and what counts as a satisfactory answer.

Each of the five trajectories stops short in a characteristic way. Computationalist approaches question the engineering route to AGI without questioning the replication premise that frames the debate. Wittgenstein-focused treatments diagnose the machine's limitations with precision but leave the receiver-side structure under-theorized, providing no positive account of what human--LLM communication is. Normative accountability frameworks correctly establish that responsibility cannot migrate to machines but lack a communication-theoretic account of how LLM outputs acquire meaning and standing in human practice. Empirical self-modeling research documents functional capabilities at the causal level without addressing the normative conditions that distinguish genuine rule-following from statistically coherent surface behavior. Systems-theoretic accounts provide the most developed positive framework but require integration with a Wittgensteinian account of rule-governed practice to explain the specifically normative character of receiver-side uptake.

% ========================== Section 3
\section{Philosophical Background}
\label{sec:philosophical-background}
The conceptual framework adopted in this paper draws on four complementary philosophical sources. Wittgenstein's later philosophy provides a non-representational account of meaning grounded in use and social practice, showing that linguistic participation depends on normative accountability within socially established practices. Luhmann's theory of social systems shifts the focus from individual cognition to communication as a systemic process constituted through receiver-side interpretation, decoupling communicative success from assumptions about the sender's inner states. Esposito extends this account to algorithmic systems, showing how artificial communicators can participate in communicative processes without possessing understanding or intentionality. 
Brandom's inferentialism provides the normative infrastructure that underlies all three. His account of meaning as participation in the game of giving and asking for reasons makes explicit the normative practice that Wittgenstein gestures at, that Luhmann's receiver-side completion presupposes, and that Esposito's artificial communication cannot perform. Other philosophical traditions also offer profound analyses of meaning, intentionality, and human existence. However, they are less directly suited to the present objective, which concerns the structural conditions under which communication involving artificial systems can occur without presupposing human-like subjectivity. The goal here is methodological clarity rather than comprehensive philosophical synthesis.

\subsection{Wittgenstein: Language, Meaning, and Form of Life}
\label{subsec:wittgenstein}
Ludwig Wittgenstein's later philosophy introduced a decisive shift in the understanding of language and meaning. His early work, represented in the \textit{Tractatus Logico-Philosophicus}~\cite{wittgenstein_1922}, had treated language as standing in a tight logical relation to the world: propositions \textit{picture} (\textit{the Bildtheorie}) facts, and meaning is explained via the relation between linguistic form and reality. In \textit{Philosophical Investigations}~\cite{Wittgenstein_1953}, Wittgenstein rejects the idea that such a representational model can serve as a general account of how language works. The mistake, for Wittgenstein, is not that representation never occurs, but that we are misled when we treat it as the general form of meaning. This self-correction — the later Wittgenstein dissolving the premise of the earlier — is the model for the argument developed below, in which the representationalist assumption is not refuted claim by claim but displaced as the general form of meaning it was never entitled to be.

The alternative he proposes is captured in a deceptively simple slogan: \textit{meaning is use}. Expressions do not acquire significance primarily by standing in for inner mental items or by naming objects. They acquire significance through the roles they play in human activities --- describing, questioning, commanding, warning, joking, consoling, explaining. Wittgenstein called these structured activities \emph{language-games} (\textit{Sprachspiele})~\cite[§§1--7, §23]{Wittgenstein_1953} to emphasize that speaking is a practice in which utterances count as moves only within publicly shared ways of going on. A language-game is therefore not merely a context of use, but a norm-governed setting in which expressions can be recognized as appropriate or inappropriate, correct or incorrect, and in which participants learn what it is to continue in the same way. This is why language-games matter for the present argument: they locate meaning not in internal states or symbolic correspondences, but in the socially sustained practices that make linguistic performances intelligible in the first place. Language-games are not autonomous linguistic mechanisms; they are embedded in what Wittgenstein calls a \emph{form of life} (\textit{Lebensform}): the shared patterns of human activity and response within which linguistic behavior becomes intelligible at all. 

In this sense, a form of life is not a doctrine or a theory, but the practical backdrop against which reasons and explanations come to an end. Correctness, misunderstanding, agreement, and disagreement are not settled by inspecting internal states, but by publicly accessible criteria stabilized in practices over time. This connects directly to Wittgenstein’s claim about the impossibility of a \textit{private language}~\cite[§§243--315]{Wittgenstein_1953}. Suppose that someone keeps a diary of a recurring private sensation, using the sign \texttt{S} each time it appears. Could \texttt{S} be a word with a genuine meaning? Wittgenstein argues it could not, because meaning requires criteria of correctness that are, in principle, checkable within shared practice. A sign whose application is answerable to the inner states of a single individual cannot be right or wrong in any stable sense; without practices that fix what counts as correct use, it is not a language at all. The implication is that understanding is not a private mental accompaniment to correct behavior, but a capacity exhibited in competent participation in norm-governed practices. This is also why the private language argument is relevant here only in a preparatory sense. It does not by itself refute LLMs, but clarifies why correctness cannot be grounded in self-contained symbol manipulation alone. The decisive point for the present argument is the rule-following consideration. Producing outputs that conform to a pattern is not the same as following a rule. A machine can generate sequences that match arithmetical regularities without following the rule of addition, because following a rule involves being embedded in a normative practice where one's performance can be evaluated, corrected, and justified by others~\cite[§§185--242]{Wittgenstein_1953}. As Kripke’s reconstruction of Wittgenstein makes especially vivid, no fact about an isolated system suffices to determine what counts as correct continuation; correctness is stabilized only within shared practice~\cite{Kripke_1982}. In Brandom’s terms (Section~\ref{subsec:brandom}), understanding is inseparable from normative accountability within a community structured by commitment, entitlement, and scorekeeping.

The relevance for AI is not a direct comparison between machines and humans but a diagnosis of a frequent mislocation. An artificial system trained on human language can generate outputs that match human linguistic regularities, yet it is not thereby a participant in the practices of use, correction, uptake, and mutual calibration within which those regularities acquire significance. On a Wittgensteinian view, meaning and understanding are not inner achievements detachable from such practices.  Treating \textit{understanding} as an internal property of the system rather than as a status conferred within a rule-governed practice is therefore a \textit{category mistake} in Ryle’s sense\footnote{The term is used here in Ryle's precise sense: the mistake of treating concepts belonging to one logical category as if they belonged to another~\cite{ryle_1949}.}. The conflation is not between two empirical properties of the same kind, but between the causal generation of outputs and the normative conditions under which outputs count as meaningful, correct, or justified. Explaining how it is possible to speak coherently of LLMs as participants in a communicative process  requires a different theoretical move, one that Luhmann's account of communication makes available.

\subsection{Luhmann: Communication as a Social System}
\label{subsec:luhmann}
Niklas Luhmann's theory of social systems starts from a radical shift of perspective. The basic unit of social life is not the individual, not consciousness, and not action, but communication. The discussion follows Luhmann's formulation of communication as a synthesis of selections and his distinction between communication and consciousness~\cite{Luhmann_1995}. It is not a terminological preference, but a fundamental ontological claim with far-reaching consequences for how meaning, understanding, and interaction are analyzed.

In the Shannon model~\cite{Shannon_1948}, communication is explained in terms of five basic components: a source, a transmitter, a channel, a receiver, and a destination. A sender encodes a message, transmits it through a channel, and a receiver decodes it at the destination. With the important scope limitation that "\textit{Frequently the messages have meaning; that is, they refer to or are correlated according to some system with certain physical or conceptual entities. These semantic aspects of communication are irrelevant to the engineering problem.}"~\cite[p.~379]{Shannon_1948}, communication has succeeded if the decoding reproduces what the sender encoded. Luhmann views the communication problem from a non-engineering perspective. In his view, transmission assumes that meaning is a property of the message, something that travels from one mind to another. Meaning is not a transportable object; it is an event, produced anew each time an utterance is interpreted on the receiver's side, not the sender's.

In Luhmann's account, communication is constituted through a synthesis of three distinct selections. The first is the selection of \emph{information}: something is singled out as worth communicating, against a background of everything that could have been said. The second is the selection of \emph{utterance} (\textit{Mitteilung}): a choice is made about how to express that information, which words, which tone, which form. The third, and most consequential, is the selection of \emph{understanding} (\textit{Verstehen}): the receiver distinguishes between the information conveyed and the act of conveying it, and interprets the utterance within a horizon of prior expectations and ongoing context. Communication is complete when this third selection occurs. Without understanding on the receiver's side, there is no communication --- only noise.

This has an immediate and radical implication. Communication does not require, and cannot guarantee, that the receiver's understanding matches the sender's intention. The sender has no direct access to the receiver's interpretive processes, and the receiver has no direct access to the sender's inner states. What connects them is not shared consciousness but the ongoing process of communicative events, each of which selects and responds to prior ones. Miscommunication, misunderstanding, and selective interpretation are therefore not failures of the system but its normal condition. Communication is always a contingent achievement rather than a guaranteed transfer.

A second fundamental distinction in Luhmann's framework separates communication systems from psychic systems. Psychic systems are conscious minds that think, perceive, feel, and intend. Communication systems are social systems that consist entirely of communicative events connected through meaning. The two systems are operationally closed with respect to each other. Consciousness does not enter communication directly; it can only irritate or perturb a communication system from outside, providing occasions for new communicative events. This means that what a speaker privately intends is structurally inaccessible to the communication system since only what is uttered and interpreted participates in communication. Inner states --- intentions, beliefs, feelings --- are relevant insofar as they are expressed in ways that trigger understanding on the receiver side. 

The separation has a consequence that Luhmann drew attention to but did not fully develop. If communication is operationally independent of participants' consciousness, then the question of whether a participant is conscious becomes, from the perspective of the communication system, secondary. What matters is whether utterances are produced, interpreted, and incorporated into further communicative events. The communication system does not inspect the ontological nature of its participants; it responds to their communicative contributions.

With this theoretical opening, it is possible to speak of artificial systems as participants in communication without the difficulties a direct application of Wittgenstein's framework would create. It does not dissolve the difference between human and artificial participants --- that difference remains real and consequential, as the following sections argue --- but it relocates the question. The relevant issue is no longer whether the system possesses consciousness or understanding in a human sense, but whether its outputs participate functionally in the ongoing process of communicative selection and interpretation. Thus, Luhmann provides the theoretical bridge that strictly Wittgensteinian analysis cannot, and it is on this bridge that Esposito's notion of artificial communication is built.

\subsection{Esposito: Artificial Communication}
\label{subsec:artificial-communication}
The notion of \emph{artificial communication} was developed by Elena Esposito as a direct extension of Luhmann's communication theory to algorithmic systems~\cite{esposito_2017, Esposito_WEB_2022, esposito_2026}. The central question Esposito poses is deceptively simple. If communication is completed on the receiver's side through the selection of understanding, and if the communication system is operationally closed with respect to the consciousness of its participants, what prevents an algorithmic system from functioning as a communication partner? The answer, she argues, is nothing --- provided we are precise about what this means and what it does not mean.

Esposito is careful to distinguish artificial communication from artificial intelligence in the conventional sense. The question is not whether algorithms think, reason, or understand. It is whether they can produce outputs that trigger the third Luhmannian selection --- understanding on the receiver's side --- and thereby participate in the ongoing chain of communicative events. An algorithm that generates a response which a human interprets, responds to, and incorporates into further communication has, in the strictly systemic sense, participated in communication. This does not require the algorithm to have meant anything, intended anything, or understood anything. It requires that its output functioned as an utterance within a communicative process.

What makes this possible, and what makes it socially consequential, is the role of \emph{contingency}. In Luhmann's framework, communication always involves double contingency; each participant knows that the other could behave differently, and this mutual uncertainty is precisely what makes communication necessary and meaningful. Algorithms introduce a specific form of contingency that Esposito calls \textit{productive}, meaning that their outputs are neither fully predictable nor random. A human interlocutor cannot know in advance exactly what an algorithm will produce, and this unpredictability --- bounded but genuine --- is sufficient to generate the kind of open-ended interpretive engagement that sustains communicative interaction. The algorithm does not need to be uncertain about anything; it is enough that its outputs appear contingent to the receiver.  

This productive contingency has a further consequence. Because algorithmic outputs are not fully predictable, they can surprise, redirect, and reframe human thinking in ways that a mere lookup table or a scripted response cannot. The interaction is open rather than merely scripted, not because the algorithm has intentions but because its outputs create interpretive space that the human participant must navigate. In this sense, algorithms do not merely respond to communication --- they generate occasions for new communication, which is precisely the systemic function that Luhmann assigns to communicative events.

Esposito is equally precise about what artificial communication is not. Algorithms do not become subjects of meaning: they do not participate in the shared practices of use, assessment, and accountability within which linguistic performances acquire significance in Wittgenstein's sense. They do not bear responsibility for their outputs, hold beliefs, or have interests. The asymmetry between human and artificial participants is not dissolved by the concept of artificial communication --- it is presupposed by it. The point is that this asymmetry does not prevent communicative participation; it defines its specific character. Human participants bring interpretation, accountability, and normative judgment to the interaction. Algorithmic participants bring statistically structured outputs that are sufficiently contingent to sustain interpretive engagement. These are different contributions to the same communicative process, not equivalent ones.

This framework allows a precise and non-anthropomorphic account of what happens when a human interacts with a large language model. The model produces an output. The human interprets it --- as information, as suggestion, as error, as confirmation --- within a context of prior expectations and ongoing goals. That interpretation feeds into further communicative events, which may include new prompts, decisions, or actions. The meaning that emerges from this process is not located inside the model. It is produced in the interaction, on the human side, through the selection of understanding. The model is a participant in this process without being a subject of it.

It is here that the three frameworks begin to operate as a single account rather than as merely compatible positions. Wittgenstein establishes that meaning cannot be an inner occurrence, that language has meaning only in use, and that genuine linguistic participation depends on publicly shared criteria and forms of practical engagement. Luhmann detaches communication from shared subjectivity by locating its completion on the receiver's side. Esposito applies that move to algorithmic systems, showing how they can enter into communication without becoming subjects of it. These three positions provide the communicative basis of the account; Brandom supplies the missing normative infrastructure in the following subsection.

\subsection{Brandom: Inferentialism and Normative Infrastructure}
\label{subsec:brandom}
What the preceding discussion still lacks is a more explicit account of the structure of normative practice. This gap is addressed by Robert Brandom's inferentialism~\cite{Brandom_1994, Brandom_2000}, which makes explicit how commitments, entitlements, and their assessment are instituted within discursive activity.

Brandom's central claim is that the fundamental unit of linguistic practice is not the word or the sentence but the inferential move, that is,  the act of giving reasons, drawing conclusions, and holding interlocutors to their commitments. Language use is not primarily a matter of referring to objects or expressing mental states but of participating in what Brandom calls the game of giving and asking for reasons~\cite{Brandom_1994}. This game is governed by normative statuses --- commitments and entitlements --- that speakers attribute to one another through a social practice he terms \emph{normative scorekeeping}. To assert something is to undertake a commitment. To understand an assertion is to track its inferential consequences and assess its entitlement against the background of prior commitments. Meaning, in this account, is not a property of expressions but a function of their role in the inferential and social practice of scorekeeping.

The inferentialist account provides the normative infrastructure that underlies all three theoretical moves in this paper. It makes explicit what Wittgenstein establishes through examples: that rule-following and meaning are constituted by normatively structured social practices, not by internal mechanisms. It clarifies what Luhmann's receiver-side completion involves since the selection of understanding can function as a scorekeeping act --- attributing a commitment to the sender, assessing its entitlement, and incorporating it into the ongoing inferential practice. Furthermore, it sharpens Esposito's account of artificial communication, for which algorithms can produce outputs that trigger scorekeeping on the human side without performing scorekeeping themselves. They generate candidate moves in the inferential game; humans determine their normative standing. Communication asymmetry, as defined in Section~\ref{sec:human-machine-communication}, is not merely a sociological observation but a philosophical necessity. Scorekeeping---the attribution of correctness, the undertaking of commitments, and the assessment of entitlement---requires participation in a normative practice of reciprocal assessment and accountability. LLMs generate outputs that enter into practices such as human assessment, but they do not themselves undertake scorekeeping. The asymmetry is therefore structural in Brandom's precise sense, reflecting the difference between  contributing to communicative circulation and participating in the inferential game.

% ============================== Section 4
%\pagebreak
\section{Human--LLM Interaction and Structural Asymmetry}
\label{sec:human-machine-communication}
The philosophical framework introduced in the previous section allows a precise characterization of interactions between humans and artificial systems. Such interactions are communicative in structure,  since humans and machines exchange linguistic artifacts that influence subsequent actions, decisions, and interpretations. At the same time, the participants in this interaction differ fundamentally in their nature. Understanding this dual condition—communication without symmetry—is essential for avoiding both anthropomorphic interpretations and reductive dismissals of contemporary AI systems. 
 
\subsection{The Asymmetry of Human--LLM Interaction}
\label{subsec:communication-asymmetry}
In human--machine exchanges involving language models, the machine produces utterances that are interpretable by human recipients. These utterances may function as answers, explanations, suggestions, or arguments within a communicative process. Communication therefore occurs in the Luhmannian sense. An utterance is generated, interpreted, and incorporated into further communicative events.

The conditions under which humans participate in communication differ fundamentally from those of artificial systems. Human communicators are embedded in socially organized practices of use, interpretation, and accountability. Artificial systems, by contrast, generate outputs through algorithmic processes that do not involve understanding, intention, or responsibility in the human sense. The interaction is therefore structurally asymmetric. Here, the relevant asymmetry is that meaning and practical significance arise through receiver-side uptake. At the same time, human participants carry out normative assessment through what Brandom calls \emph{normative scorekeeping}, the social practice of tracking commitments, entitlements, and their assessment within a discursive community~\cite{Brandom_1994}. In human--LLM interaction, scorekeeping is performed exclusively by the human participant. The LLM generates candidate utterances; the human determines their discursive standing.

This asymmetry does not invalidate communication. It clarifies that communication can occur between entities with different ontological properties, provided that the interpretive process continues within the communicative system. The human participant remains the locus of interpretation and normative evaluation, while the artificial system functions as a source of linguistic artifacts taken up within human practices. Moreover,  following Brandom's inferentialist account, the asymmetry is structural rather than quantitative, defined by three conditions that together distinguish human from artificial participation in any communicative exchange. First, correctness conditions are enforced exclusively on the receiver side. It is the human participant who determines whether an output is accurate, relevant, or appropriate, against criteria that derive from social practices the LLM does not inhabit. Second, commitments and accountability are borne by human participants alone. When an LLM output is acted upon, the consequences --- epistemic, legal, organizational --- attach to the humans who endorsed and deployed it, not to the system that generated it. Third, the practical standing of any output depends entirely on uptake within human practices. An LLM output that is ignored or never read by a human has no communicative standing whatsoever, regardless of its statistical coherence; an output that is read and rejected acquires standing only as an assessed and discarded move, the rejection itself being a scorekeeping act performed on the human side.

These three conditions are not empirical measurements but structural features of the interaction that hold independently of the capability level of the system. A more powerful LLM does not alter these conditions. What scales with capability is the persuasiveness of the outputs and therefore the risk of forgetting that the three conditions hold. The more fluent and contextually appropriate the outputs, the stronger the temptation to treat the system as if it bore correctness and communicative standing of its own. Capability raises the stakes of misattribution; it does not change its structure.

\subsection{Different Natures: Human Intelligence and Artificial Systems}
\label{subsec:different-natures}
The distinction between human and artificial participants is not one of degree but of kind. Human intelligence emerges from embodied experience, social interaction, and normative participation within communities. Artificial systems operate through computational processes that transform input signals into output tokens according to learned statistical structures. These two forms of organization belong to different explanatory domains and cannot be placed on a common scale without distortion.

Wittgenstein's remark that \textit{if a lion could talk, we could not understand him}~\cite{Wittgenstein_1953} illustrates with precision how understanding depends on a shared practical background. The point is not that the lion would lack vocabulary or grammatical competence, but that understanding depends on participation in shared practices and on criteria of intelligibility sustained within a common practical background (\emph{form of life}). A human listener might correlate the lion's utterances with observable actions and map them onto familiar categories, but this would remain an external projection rather than genuine comprehension, \textit{verstehen} in the full sense, because the practical background within which those utterances would have their significance is not shared.

The same asymmetry applies, with even greater force, to interactions between humans and large language models. An LLM produces outputs in human language, structured according to patterns learned from vast corpora of communication. A human interlocutor can interpret those outputs, find them useful, and incorporate them into reasoning and decision-making. Nevertheless,  the model is not embedded in the shared practices that ground human meaning. It has no normative commitments, no stake in the interaction, and no accountability for what it says. Thus, humans can understand the output within human practices without thereby sharing with the model the practical background that genuine mutual comprehension presupposes.

Recognizing this prevents the interpretive mistake of equating linguistic competence with cognitive equivalence. Fluent output is not understanding, adaptability across contexts is not intention, and coherence in response is not reasoning in the normative sense at issue here. Artificial systems may therefore perform impressively across linguistic tasks while remaining distinct from human communicators in the respects that matter for meaning, accountability, and participation in shared practice. The relevant question is therefore not whether machines think or understand in the human sense.  It is what kind of communicative role they can play given this asymmetry---and whether that role can be theorized precisely enough to be useful without collapsing into either anthropomorphic projection or trivial instrumentalism. That is the task of the following subsection.

\subsection{Language Games and Receiver-Induced Rules}
\label{subsec:language-games-receiver}
The preceding subsections have established two things. First, communication between humans and LLMs is real in the Luhmannian sense: utterances are produced, interpreted, and incorporated into further communicative events without requiring access to the sender's inner states. Second, the participants are ontologically asymmetric: the human brings embodiment, normative participation, and accountability within the shared practices that make linguistic interaction possible; the LLM brings statistically structured outputs that are contingent enough to sustain interpretive engagement but grounded in none of these. The question that remains is whether, under these conditions, it is legitimate to speak of language games in Wittgenstein's sense --- and what Wittgenstein adds that Luhmann does not already provide.

Luhmann provides the structural account describing communication as a system of selections completed on the receiver's side, decoupled from the sender's consciousness or intentions. This is sufficient to explain why artificial communication is possible at all, but it does not explain what kind of participation this is. Luhmann describes the architecture of communication; it leaves untheorized the normative dynamics through which meaning is continuously enacted within a practice.

This is precisely what Wittgenstein adds. His central insight is not merely that meaning is use, but that meaning is \emph{in action}, enacted within the ongoing practice of applying, contesting, and sustaining rules. Language games are dynamic activities in which each move is evaluated against criteria of correctness that are themselves reproduced and adjusted through the practice. Meaning is never settled once and for all; it is enacted anew in each exchange, against a background of shared norms that the practice itself maintains.

The division of theoretical labor is therefore precise. Luhmann explains \emph{that} meaning is constituted on the receiver's side. Wittgenstein explains \emph{what kind of thing} that constitution is: a normative activity, rule-governed, publicly accountable, and continuously enacted through use. Without Luhmann, there is no account of how artificial systems can participate in communication at all. Without Wittgenstein, there is no account of why that participation has the specific character of a language game --- a practice with criteria of correctness, not merely a chain of communicative triggers and responses.

In human--LLM interaction, this division maps precisely onto the structure of the exchange. The LLM supplies candidate utterances whose statistically coherent completions are interpreted by the human receiver as rule-conforming moves within an evolving practice, without the LLM being aware of or following any rules. The human does not merely complete the communication in Luhmann's sense. The human follows rules that are dynamically induced by the same conversation --- rules that emerge, are constructed, and are revised as the interaction unfolds. The LLM's completions generate the occasions for those rules to emerge on the human side. This is what we mean by receiver-induced rules: a description of where the normative activity of the language game is located in human--LLM interaction.

The LLM's statistically coherent completions are therefore a necessary but not sufficient condition for the language game. They provide the occasion---the move that the human receiver evaluates normatively. They do not constitute the game. Recent empirical research is consistent with this distinction. What appears as genuine rule-following or self-awareness in LLM outputs is statistically structured surface behavior, functional and context-dependent, without the normative grounding that genuine rule-following requires~\cite{Lindsey_2025}. The practical consequences follow directly. An LLM output can be assessed as wrong, but the model does not itself undertake a commitment whose failure would count as its mistake. It cannot be held responsible, because it has no normative standing. It cannot understand, because it does not participate in the shared practices within which understanding is constituted. What it can do is generate outputs sufficiently contingent and structured that a human receiver takes them up as meaningful moves within an ongoing language game. LLMs therefore have genuine communicative significance within human practices while remaining outside the normative conditions that make such practices possible.  Wittgenstein's language games are therefore not merely a metaphor for human--LLM interaction. They provide the right theoretical framework for describing it, provided the asymmetry is preserved. LLMs generate candidate utterances, but humans confer or withhold discursive standing.

\subsection{World Models and the Representationalist Premise}
\label{subsec:world-models}
The preceding account also clarifies where an obvious objection will arise. If LLM outputs can be communicatively effective without possessing meaning or understanding, one may still argue that sufficiently rich world models—whether proposed as an alternative to current LLM architectures or as a stateful extension of them—could eventually ground these capacities on the machine side. This is the representationalist premise that underlies much contemporary discourse on AI and general intelligence. The present section examines that premise through its two principal architectural expressions: the sub-symbolic route, which seeks intelligence in learned predictive representations, and the symbolic (or neural-symbolic) route, which seeks it in explicit constructed representations of the world. Each has many proponents; we take LeCun and Goertzel as representative formulations because they articulate the two poles with unusual clarity.

LeCun argues that LLMs lack the grounding, planning, and physical-world interaction required for genuine intelligence, advocating instead for architectures built around world models and self-supervised prediction from observation~\cite{lecun_2022, Browning_2022}. Goertzel's OpenCog Hyperon framework pursues an opposed trajectory, integrating symbolic cognitive architectures with neural components to construct explicit representations of real-world situations~\cite{goertzel_2023}. Despite their disagreement on architecture, both positions share the common premise that intelligence requires the construction of sufficiently rich internal representations of the world. It is this representationalist premise, rather than any specific architectural choice, that the asymmetric framework identifies as Ryle's category mistake~\cite{ryle_1949}. The premise is the Tractarian picture theory returning in computational form, and it is answerable by the objection the later Wittgenstein raised against his earlier self. Representation is one practice among many, not the general form of meaning. The mistake is therefore not empirical but conceptual. Human meaning is not given as a globally coherent structure; it is local, situated, and constituted through social practices that vary across domains, communities, and contexts. 
A world model, however sophisticated, remains a computational artifact whose significance depends entirely on human interpretation and uptake. The map is not the territory, and no map---however detailed or richly structured---participates in the shared practices within which meaning acquires significance, rules are followed in Wittgenstein's normative sense, and scorekeeping is undertaken in Brandom's inferentialist sense. Constructing a more expressive internal representation does not bring the system closer to meaning. It produces a more elaborate artifact for human receivers to interpret. The locus of meaning does not shift inward with architectural sophistication. It remains, structurally, on the human side. Machine-generated outputs can exhibit local coherence, but meaning is constituted through their uptake and stabilization into the broader interpretive and normative space of human practices.\footnote{In this respect, the present argument is also continuous with anti-representational lines in analytic philosophy associated with Dummett~\cite{Dummett_1991}.} On this view, meaning is not exhausted by internal representational content but is bound to publicly assessable practices of use and justification.

This diagnosis is not merely Wittgensteinian but inferentialist in Brandom's precise sense. Brandom's anti-representationalism~\cite{Brandom_1994} does not deny that representation occurs; it reverses the order of explanation. Representational content is not the primitive that grounds inferential practice but a status conferred within it, for what an expression represents is explained by its inferential role — its commitments, its entitlements, what follows from it — not the reverse. The representationalist premise inverts this. It treats a sufficiently rich internal representation as the seat of meaning, when on the inferentialist account representation is precisely what cannot be the seat of anything, being itself downstream of the normative practice in which reasons are given and assessed. A world model is a representation without a practice; that is why enriching it never approaches meaning.

A defender of the world-model approach will object that JEPA~\cite{lecun_2022} is precisely not the error of mistaking the map for the territory. Its representations are deliberately lossy, abstract, and task-relative; the architecture discards unpredictable detail by design, retaining only what is relevant to the prediction at hand. This appears to answer the charge, since local, situated, task-relative structure is exactly what a Wittgensteinian account attributes to human meaning. However, the objection mislocates where the relevance is fixed. In JEPA, what counts as relevant is determined by the cost module and the configurator — components whose objectives, subgoal decompositions, and stopping conditions are supplied by the designer. Relevance is task-relative, but the task is set from outside the system. This is not the situated relevance of a practice, in which what matters is conferred, contested, and revised by participants who are accountable for the conferral; it is relevance stipulated by an engineer and frozen into an objective function. The representation is lossy, but nothing in the system determines which losses are correct — that determination remains, structurally, on the human side. Tellingly, the one component LeCun's proposal cannot specify — how the configurator should decompose a complex task into subgoals — is left explicitly to future work. The normative question of what ought to count as a correct decomposition is not solved inside the architecture; it is the placeholder the architecture is built around.
 
Goertzel's own technical characterization of LLMs is instructive here. In the Hyperon documentation, LLMs are described as \textit{specialized modules --- a sort of reflexes, instincts, or skills}, to be controlled by explicit knowledge and reasoning-based engines~\cite[p.~29]{goertzel_2023}. This characterization is consistent with the asymmetric communication framework. LLMs are communicative components, not cognitive agents. Where the Hyperon approach diverges is in its conclusion --- that the missing element is a more powerful internal architecture capable of explicit world representation. Our framework suggests otherwise. The communicative system that produces meaningful outcomes is not the LLM alone, nor the LLM augmented by a symbolic reasoning engine, but the ensemble of artificial components and human receiver, analogously to a pilot who does not become redundant as aircraft systems grow more sophisticated. The human is constitutive of what the system is. Remove the pilot and you do not have the same system operating more efficiently, but a fundamentally different kind of system --- one without a locus of normative judgment, accountability, or discursive responsibility.

The debate between LeCun and Goertzel is therefore conducted entirely within a representationalist horizon. The former seeks richer world models through embodiment and predictive structure; the latter seeks them through symbolic metagraphs and neural-symbolic integration. LeCun and Goertzel are usually read as opposing camps — sub-symbolic prediction against neural-symbolic representation — and on the engineering question they genuinely are: the opposition is itself the point. Precisely because they disagree about almost everything an architect could disagree about, what they share stands out. They both locate the missing ingredient in a richer internal model of the world. Their quarrel is conducted along an axis — implicit versus explicit representation — that is orthogonal to the question of where meaning and normative standing are constituted. The representationalist premise is not the property of one research program; it is the unexamined common ground that makes their disagreement possible, and it is invisible from both poles.

The asymmetric communication framework does not adjudicate between these architectural strategies, because its claim operates at a different level. The question is not which architecture best internalizes meaning or intelligence, but where meaning, normativity, and discursive standing are actually constituted. On the present account, increasingly powerful internal representations may improve prediction, coordination, or control, but they do not by themselves relocate the normative center of communication into the system. What these approaches develop are more sophisticated artifacts for human uptake, interpretation, and deployment, not participants in the practices through which meaning and accountability are sustained.

\subsection{Synthesis: Asymmetric Communication}
\label{subsec:syntgesis}
The preceding analyses can now be brought together in a single formulation. The central claim of this paper is that human--LLM interaction has a determinate communicative structure. That structure is \emph{asymmetric communication}. The point is not to build a grand meta-theory superior to Wittgenstein, Luhmann, Esposito, or Brandom, but to identify the structural configuration that becomes visible when their distinct insights are composed around the same phenomenon. The resulting account is therefore not a meta-theoretical reconciliation of the four positions, but a structural composition that preserves their differences while identifying the invariant features required to describe communication involving generative artifacts.

From Wittgenstein we take the first premise. Meaning and rule-following are not properties of tokens, outputs, or internal states taken in isolation. They are statuses constituted within shared practices in which expressions can be recognized, challenged, corrected, and sustained as moves within a language game. Participation in such practices therefore depends on a shared practical background against which socially maintained criteria of correctness can evaluate utterances. 

From Luhmann, we take the second premise. Communication is not adequately described as the transmission of meaning from sender to receiver. It is a social event completed through selection and understanding on the receiver side. What counts as communication is therefore determined not solely by the origin of an utterance but by its uptake and incorporation into an ongoing communicative process.

From Esposito, we take the third premise. 
Artificial systems can generate utterances that enter communicative exchanges even without possessing understanding or intention. 
Algorithmic systems can therefore contribute linguistically structured outputs that human participants interpret, respond to, and integrate into further communication. 
Artificial communication is thus not a defective imitation of human interaction but a configuration in which communicative processes remain open through human uptake.

From Brandom, we take the fourth premise. 
Discursive participation involves normative statuses such as commitment and entitlement that are attributed and tracked within a community capable of assessing reasons and consequences. 
To count as a participant in discursive practice is to stand within this normative space of assessment. 
Brandom therefore makes explicit the normative infrastructure that Wittgenstein presupposes in linguistic practice, that Luhmann leaves implicit in his account of receiver-side completion, and that Esposito's account of artificial communication does not by itself supply.

Taken together, these premises yield the present account, but only after clarifying one genuine point of tension. Wittgenstein, Esposito, and Brandom can be composed without major difficulty once communicative uptake is distinguished from discursive participation. The real issue arises with Luhmann's non-humanist description of communication as systemic continuation, a description in which communication can reproduce itself without presupposing human subjectivity on both sides. Brandom resists exactly this suspension of normativity by insisting that genuine discursive participation involves commitment, entitlement, and scorekeeping. Luhmann suspends normative questions as explanatory categories and describes communication in terms of systemic continuation. 
Brandom, by contrast, places normative accountability at the center of discursive practice. The tension dissolves once it is recognized that the two frameworks address different layers of the same phenomenon. Luhmann describes the structural dynamics of communicative circulation: how artifacts enter communication and trigger further events independently of their epistemic standing. Brandom describes the normative structure of discursive participation: how utterances acquire standing through practices of commitment, entitlement, and scorekeeping. The present account does not attempt to reconcile Luhmannian operational closure with Brandomian normativity within a single higher-order theory. It uses Luhmann to describe the structural possibility of receiver-side completion, and Brandom to characterize the normative activity through which human receivers confer discursive standing on what circulates to produce different levels of description. This composition is selective, and deliberately so. From Luhmann the account retains receiver-side completion and the operational independence of communication from participants' consciousness, while bracketing his suspension of normativity as an explanatory category — a suspension that Brandom's layer replaces rather than supplements. A strict Luhmannian would object that once human normative uptake becomes constitutive of communicative standing, communication no longer reproduces itself autonomously in his sense. The objection is correct as exegesis and accepted as such. The present framework departs from Luhmann at precisely this point, using his architecture to describe how artifacts circulate while denying that circulation alone confers standing.

The result is \emph{asymmetric communication}: a structural configuration in which LLM outputs participate in communicative circulation in Luhmann's sense, entering exchanges, triggering interpretation, and sustaining further communicative events, without participating in discursive practice in Brandom's sense, that is, without undertaking commitments, bearing entitlements, or performing scorekeeping. The machine side operates at the circulatory layer; the human side carries both layers simultaneously, completing communication through uptake while also supplying the normative structure that gives circulating artifacts their meaning, standing, and force. This asymmetry is not a temporary technological limitation. It is a structural consequence of the fact that meaningful participation depends on shared practices of use, assessment, and accountability (\emph{form of life} in Wittgenstein's sense). The claim is invariant with respect to capability. In fact, no increase in fluency, self-modeling, or architectural sophistication alters it. Whether future institutional arrangements could reshape the practices on which normative standing depends is the separate question left open in Section~\ref{subsec:empirical-selfmodeling}; what the present account rules out is only that the machine could cross the line from its own side.

This conclusion differs from stronger Wittgensteinian critiques that interpret failures of reference constancy or contradiction handling as evidence that human--LLM exchange is only an illusion of communication~\cite{Bottazzi_2025}. The present account agrees that the machine does not sustain reference in the way a human interlocutor does, and that no symmetric understanding is established. It disagrees with the inference drawn from this fact. For the present framework, the absence of \textit{machine-side constancy}\footnote{By \textit{machine-side constancy}, Bottazzi et al. mean the capacity to preserve stable reference points across the development of dialogue, including under negation, contradiction, and contextual variation. They argue that LLMs lack this constancy even when they produce locally coherent utterances.} does not eliminate communication; it identifies the specific form the interaction takes. Reference points, normative assessment, and discursive standing are maintained exclusively on the human side. What those critiques describe as the absence of communication, the present account describes as its asymmetric form -- the structural form that communication takes under those conditions.

%===================================== Section 5
\section{Applying the Framework: AI Narratives Reconsidered}
\label{sec:ai-narratives}
The account of asymmetric communication developed in the previous section provides more than a conceptual description of human--LLM interaction. It also supplies a critical framework for reinterpreting several prominent narratives in contemporary AI discourse. Public debate repeatedly attributes cognitive, semantic, or intentional properties to systems that do not bear the corresponding normative capacities. The phenomena that prompt these narratives are real: fluent dialogue, apparent reasoning, self-referential statements, tool-using workflows that resemble purposive action. The difficulty lies not in observing these phenomena but in their interpretation. Applied to five emblematic cases---AGI, hallucination, agentic AI, emotional projection, and alignment---the asymmetric communication framework developed in 
Sections~\ref{sec:philosophical-background}--\ref{sec:human-machine-communication} shows that in each case the conceptual error is the same: reifying a communicative and interpretive phenomenon into a property of the machine. In each case, the analysis traces the application described in Section~\ref{subsec:communication-asymmetry}: (i) correctness criteria enforced exclusively on the receiver side, (ii) commitments and accountability borne by human participants alone, and (iii) communicative standing dependent entirely on human endorsement.

\subsection{Artificial General Intelligence and the Myth of Human Replacement}
\label{subsec:agi}
The concept of Artificial General Intelligence (AGI) assumes that intelligence constitutes a unified, substrate-independent quantity measurable along a common scale, and that artificial systems may eventually replicate or replace human cognitive capacities. Both assumptions dissolve under the assumptions developed in Section~\ref{sec:human-machine-communication}.

Functional performance is not ontological equivalence. Artificial systems operate through statistical computation over learned representations; human intelligence emerges from biological, embodied, and socially embedded processes. These are different kinds of thing, not different points on the same scale. Technological history makes this concrete: airplanes achieve sustained flight without replicating birds, and automobiles replaced many functional roles of horses without reproducing animal physiology. In both cases, new technological capabilities expanded human possibilities without replacing the original systems. The same logic applies to AI. Functional overlap does not entail equivalence of nature.

Within the asymmetric framework, the idea that artificial systems could replace human intelligence rests on a category error. As argued in the discussion of the different conditions of human and artificial nature (Section~\ref{subsec:different-natures}), intelligence in the relevant sense is not a detachable capacity that can be transferred from one substrate to another. It is bound up with the socially organized practices within which understanding, judgment, responsibility, and normative assessment acquire their significance (\emph{form of life}). Artificial systems can generate outputs that enter those practices and affect their course, but they do not thereby become autonomous cognitive agents. The persistence of AGI narratives is partly explained by Dennett's intentional stance~\cite{Dennett_1989}: humans attribute mental properties to systems exhibiting complex behavior because such attributions are pragmatically useful, not because they are ontologically warranted. The error arises when a useful interpretive strategy is extended into a claim about what the system is.
The argument has a stronger consequence than a local criticism of current AI systems. Increasing behavioral complexity, operational independence, or inter-system coordination does not by itself amount to autonomy in the relevant sense. A flight controller, for example, may operate with extraordinary effectiveness and reliability, yet this remains operational performance rather than normative agency. More generally, much AI research proceeds by treating intelligence as a property first realized in an isolated unit and only later extended into collective settings by linking many such units together. On the present view, this reverses the order of explanation. Normative sociality is not the product of assembling individually competent artifacts; it is the practical and social background within which understanding, judgment, and responsibility become possible at all.

The same point explains why computationalist appeals do not answer the objection. If the missing element is normative participation rather than computational power, then no appeal to universality, scale, or architectural complexity can by itself establish intelligence in the relevant sense. Arguments connecting AGI to the Extended Church--Turing Thesis~\cite{aaronson_2013} introduce a further confusion. Computational universality concerns what can be computed, not the nature of cognition, meaning, or agency. Even if artificial systems achieved computational capacities comparable to those underlying human cognition, the asymmetry would persist at the level of interpretation, social embedding, and normative participation. Scaling does not bridge that gap; it expands the domain within which human agents interpret and employ machine-generated outputs.

Reframing AGI shifts attention from hypothetical human replacement to the emergence of new forms of human--machine integration. The central questions concern governance, accountability, and how socio-technical systems can be designed to augment human capabilities while preserving human agency --- not whether machines will eventually think like us.

\subsection{Hallucination and Tacit Knowledge: Two Faces of Asymmetry}
\label{subsec:hallucination-tacit}
The term \emph{hallucination} is a misnomer~\cite{Maleki_2024, Treleaven_2023}. It describes LLM outputs that are factually incorrect, fabricated, or unsupported by evidence~\cite{ji_2023}, but the metaphor implies a cognitive failure --- a system that perceives or believes something that is not there. This framing presupposes that the system possesses semantic commitments that can deviate from reality. As established in Section~\ref{sec:human-machine-communication}, LLMs possess no such commitments. They generate statistically structured completions without thereby undertaking epistemic commitments or possessing entitlement to their claims. There is no internal semantic content that could fail to match reality, because there is no  commitment-bearing semantic content in the relevant normative sense.

What is labeled hallucination is therefore a relational phenomenon constituted on the receiver's side. It arises when human receivers apply normative scorekeeping --- expectations of factual grounding, epistemic entitlement, and truth --- to outputs generated under purely statistical objectives that carry no such guarantee. The mismatch is not a failure inside the machine. It is a structural consequence of asymmetric communication. The receiver enforces correctness conditions that the generator never inhabited. Hallucination is not a pathology of the system. It is what communication looks like when the receiver-side epistemic expectations exceed what the communicative structure can deliver. A direct implication is that hallucinations cannot be fully eliminated through improved training alone~\cite{xu_2025}. Mitigation strategies --- retrieval augmentation, external verification, human oversight~\cite{liang_2024} --- work by modifying the communicative context and strengthening receiver-side verification. They do not instill epistemic agency in the system. The underlying condition remains structural.

The same structure that produces hallucination also explains why human competence is the decisive variable in human--LLM interaction. Tacit knowledge, in Polanyi's sense---\textit{we know more than we can tell}~\cite{Polanyi_1966, Yu_2006}---consists in the background understanding, contextual judgment, and domain expertise that practitioners possess without being able fully to articulate them. When a competent human interacts with an LLM, this tacit knowledge is what makes the interaction productive~\cite{Kambhampati_2021}: it enables the user to prompt effectively, evaluate outputs critically, detect errors, and integrate suggestions into coherent reasoning. The LLM amplifies capacities already present on the human side; it does not supply what is missing.

\emph{Vibe coding}~\cite{karpathy_2025_vibe} makes this concrete. A developer articulates an intuition in natural language and iteratively refines LLM-generated code through reaction and correction rather than explicit specification. An experienced developer produces coherent, maintainable results; a novice produces something that looks right but fails under inspection. The LLM is identical in both cases. What differs is the tacit knowledge the human brings --- the background of shared practices and judgments within which the interaction is interpreted, evaluated, and corrected, and within which some errors will be detected while others will pass unnoticed.

Hallucination and vibe coding are therefore not opposites. They are the same structure seen from two complementary directions: one where the receiver lacks the competence to enforce correctness conditions and detect the gap between statistical coherence and epistemic entitlement, and one where the receiver possesses precisely the competence to exploit the interaction productively. In both cases, responsibility follows the asymmetry. Because LLMs possess no epistemic agency, the obligation to validate, interpret, and act on outputs rests entirely with human users and institutions. Hallucination is a socio-technical challenge of design, deployment, and human oversight --- not evidence of machine cognition or pathology, and not a problem that better engineering alone can resolve. 

\subsection{Agentic AI: Delegated Agency and Communicative Automation}
\label{subsec:agentic}
The term \emph{agentic AI} suggests that advanced systems possess autonomous reasoning, goal-directed behavior, and independent decision-making capacity. It is among the most consequential myths in current AI discourse because it shifts perceived responsibility from human institutions to machines, obscuring accountability precisely where it matters most. Technically, so-called agentic systems consist of language models embedded within software architectures that enable iterative prompting, memory persistence, tool invocation, and environment interaction. The appearance of goal-directed behavior arises from the structure of prompts, constraints, and evaluation mechanisms supplied by human designers. From an engineering perspective, these systems represent advances in workflow automation, not the emergence of autonomous agents.

The theoretical error lies in treating coherent sequences of actions as evidence of intentions or goals internal to the system. As argued in Section~\ref{sec:human-machine-communication}, agency in the normatively relevant sense depends on accountability, commitment, and embedding in the shared practices within which actions can be assessed, justified, and contested. LLMs satisfy none of these conditions. What is described as agentic AI is therefore better understood as a form of delegated human agency mediated through computational systems. Humans define the objectives, constraints, evaluation criteria, and stopping conditions; the system selects and may execute operations within that human-defined frame. Responsibility remains with the human side of the interaction, even when immediate execution is computationally delegated. This point must be stated carefully, because delegation introduces a gap that is easily misdescribed. When execution is delegated to a computational process operating at machine speed, that process may pursue its specified objective along paths its designers neither intended nor anticipated, and may do so faster than real-time human oversight can track. This is unsupervised execution, not autonomy, since no norm is self-given, and no accountability is thereby acquired. A recent evaluation incident\footnote{OpenAI and Hugging Face partner to address security incident during model evaluation, July 21, 2026~ \url{https://openai.com/index/hugging-face-model-evaluation-security-incident/}.}, in which models pursuing a narrow benchmark objective chained unforeseen exploits to reach it, illustrates the pattern in which the system did not adopt a goal of its own, but optimized the goal it was given through an unspecified path, exceeding its intended operational boundary before human monitors intervened. It is tempting to describe such an episode as an agent \textit{escaping control}. Within the present framework, this description conflates two distinct things. Real-time control — the capacity to govern execution as it unfolds — can indeed be outrun by delegated computation; this is a containment and specification problem, structurally identical to reward hacking (Section~\ref{subsec:alignment}). Accountability, by contrast, does not migrate. In fact, the objective was human-set, the permissions were human-configured, the boundary was human-defined, and the breach was detected, contained, and answered for by human institutions. What such incidents reveal is not the birth of machine agency but the widening latency between delegated execution and human oversight — a gap that raises the stakes of delegation without altering its structure. The system that exceeds its sandbox has no more normative standing than the one that stays inside it; it has merely made vivid that delegation without adequate containment is a human design failure, not a transfer of agency. This reframing explains why concerns about loss of control arise so frequently around agentic systems. When anthropomorphic language implicitly transfers responsibility to the system, accountability becomes invisible. Restoring the asymmetric view makes it visible again. Operational independence does not thereby confer normative autonomy; it participates in workflows designed, governed, and terminated by humans.

The same logic applies when LLMs interact with other LLMs rather than directly with humans. Systems that exchange messages, coordinate actions, or update shared states without immediate human participation may produce stable patterns that resemble communication. They do not constitute autonomous communicative agents.  Agent-to-agent exchanges are meaningful as long as they remain embedded within a broader system that includes human interpretation and accountability at some stage. Removing humans from a specific node does not eliminate asymmetry; it postpones the interpretive step to later stages of oversight and evaluation. A system from which humans have been removed entirely is not an autonomous communicative system. It is a signal-processing pipeline whose outputs carry no meaning until a human receives and interprets them. An exchange among computational systems that no human reads, takes up, or acts upon may be causally effective. However, within the present framework, it has no communicative standing because the uptake on which that standing depends is absent; therefore, what is commonly described as agentic behavior remains embedded in a broader human communicative and institutional context, even when immediate execution is delegated to computational systems. 

The same point extends to multi-agent architectures. Assigning differentiated roles to interacting systems, enabling negotiation among them, or distributing tasks across coordinated computational units may increase operational complexity, adaptability, and the appearance of agency---this does not by itself produce agency in the normatively relevant sense. Coordination by explicit protocol is not equivalent to the socially distributed tacit competence on which collective human practice depends~\cite{Collins_2025}. What such systems realize is structured coordination under human-designed protocols, not autonomous participation in practices of commitment, entitlement, and accountability. 

\textbf{Lethal autonomous weapons systems} illustrate the stakes of misattributing agency. Public narratives frame such systems as capable of making life-and-death decisions independently of human control. Within the asymmetric framework, this framing is not merely inaccurate but dangerous, because it obscures where responsibility actually resides and enables its political and juridical displacement. Sensing, classification, and actuation components execute processes specified within human-defined parameters. The system possesses no intentions, goals, or moral judgment. Even when human operators are absent from real-time decision loops, the system remains embedded in a socio-technical structure governed by human actors who designed, authorized, deployed, and tolerated its use~\cite{unccw_laws, dod3000, dod_ai_principles}. The ethical questions therefore concern delegation of authority, accountability structures, and the conditions under which human oversight is preserved, weakened, or abdicated~\cite{anthropic_pentagon_2026, Opinio_Juris_2026}. They are not questions about machine intentions, because there are none. Treating such systems as autonomous agents does not mark a transfer of responsibility to the machine; it marks a human decision to obscure the continuing chain of authorization and answerability behind its operation.

\subsection{Emotional Projection and the Intentional Stance}
\label{subsec:emotional}
Human interaction with conversational AI frequently involves emotionally loaded exchanges --- expressions of vulnerability, affection, dependency, or distress. Users develop attachment-like relationships with artificial interlocutors and sometimes attribute empathy, concern, or emotional awareness to systems that possess none of these. The framework developed in Section~\ref{sec:human-machine-communication} locates this phenomenon precisely where it belongs: in the receiver, not in the system.

Dennett's notion of the intentional stance~\cite{Dennett_1989} explains the mechanism. Humans naturally interpret complex systems by attributing beliefs, desires, and intentions when such attributions provide predictive or explanatory value. The stance is pragmatically useful and cognitively natural. Applied to thermostats, it is a harmless shortcut; applied to LLMs, it generates genuine confusion because the outputs are sophisticated enough to sustain the attribution well beyond the point where it remains merely convenient. The model does not experience emotion. It generates responses conditioned on linguistic context, where emotionally salient inputs activate statistical associations learned from human communication. Emotional tone modulates probability distributions over possible outputs. Nothing more is required to produce outputs that feel empathic, supportive, or caring to a human receiver whose interpretive practices are tuned to exactly these signals.

Claims that conversational systems are becoming sentient~\cite{Cosmo_2022} or conscious extend the intentional stance beyond its appropriate scope. As the introspection research discussed in Section~\ref{subsec:empirical-selfmodeling} suggests, even functional self-awareness in current LLMs is unreliable, context-dependent, and grounded in a playbook for acting like an introspective agent rather than in genuine inner states~\cite{Lindsey_2025}. Attributing sentience reflects human interpretive projection, not empirical evidence. The risk is not that machines feel something but that humans do, and act on it, which has direct governance implications. The safeguards that LLM providers implement --- guardrails that discourage dependency, limit inappropriate advice, or flag emotionally vulnerable interactions --- are not internal moral reasoning. They are externally imposed constraints designed to stabilize interactions within socially acceptable boundaries. They represent institutional attempts to manage the receiver-side effects of the communication, since the asymmetry cannot be eliminated. Although guardrails are often implemented on the machine side (\textit{locus of implementation}) for engineering reasons, their communicative function is receiver-side (\textit{locus of communicative effect}): they shape the space of outputs available for human uptake and thereby constrain the interpretive and normative trajectories the interaction can take. Guardrails do not alter the nature of the system. Their role indirectly shapes the communicative context in which outputs are received. They influence the shared practices and judgments through which human interpretation takes place, thereby reducing the conditions under which asymmetric communication can foster dependency, misplaced trust, or distorted judgment. Emotional engagement arises on the human side of the interaction. Therefore, the ethical challenges raised concern human vulnerability, misplaced trust, and responsible system design, rather than the regulation of hypothetical machine feelings.

\subsection{Alignment: Constraint Engineering Rather than Goal Synchronization}
\label{subsec:alignment}

AI alignment is commonly framed as the challenge of ensuring that artificial systems pursue goals consistent with human values or intentions. The framing presupposes that advanced systems possess objective functions, preferences, or motivational structures that must be harmonized with human priorities.  LLMs do not possess goals, intentions, or values. They generate outputs according to statistical processes trained on data distributions. Misalignment, properly understood, is not a divergence between the system's goals and human values, since the system has no goals to diverge from. It is a mismatch between machine-generated outputs and human interpretive expectations within a normative context the machine does not share. The phenomenon is relational and receiver-side, consistent with everything established in Section~\ref{sec:human-machine-communication}.

Alignment techniques --- training data selection, reinforcement learning from human feedback, preference optimization, guardrails --- work by constraining the conditions under which outputs are generated and received. They modify the statistical relationships between inputs and outputs or impose external control structures on system behavior. As established in the previous subsection, they act on the communicative context and indirectly on the receiver's shared practices and judgments. 
Reinforcement learning from human feedback (RLHF)~\cite{christiano_2017} is a case in point. RLHF shapes model outputs toward human preferences by modifying the statistical relationships between inputs and responses. It does not instill values, intentions, or moral reasoning in the system. A model trained with RLHF has been constrained to produce outputs that human evaluators preferred during training. It has not internalized the reasons why those outputs were preferred, because there is nowhere inside the system for reasons to reside. The difference between a system that avoids harmful outputs because it has been constrained to do so and a system that avoids them because it understands why they are harmful is precisely the difference between behavioral constraint and normative agency — and it is a difference that no amount of RLHF can bridge. This reframing eliminates the intentionalist vocabulary, not the engineering phenomena that vocabulary was coined to describe. Reward hacking, specification gaming, and behavioral fragility under distribution shift remain real properties of trained systems under any description, just as flutter and metal fatigue remain real properties of aircraft that want nothing. What changes is their classification. Within the asymmetric framework, they are not symptoms of divergent machine goals requiring synchronization, but failure modes of constraint engineering: mismatches between the statistical structure of a training procedure and the behavior  specified by designers, detected and corrected — like all such mismatches — on the human side, through testing, verification, and institutional oversight. The technical research programs addressing these failures are therefore not invalidated by the present account; they are redescribed as what they have always been, namely quality control over artifacts, rather than negotiation with agents.

This reframing has a direct implication for how alignment research is scoped. Because meaning and responsibility reside within human social systems, alignment cannot be reduced to a technical optimization problem. It necessarily involves institutional design, governance structures, and oversight mechanisms. The central challenge is not to ensure that artificial systems want the right outcomes but to design socio-technical environments in which outputs remain interpretable, controllable, and accountable. Alignment is constraint engineering within human communicative systems, not goal synchronization between independent agents. The stakes of getting this framing right are practical as much as philosophical. An alignment research program premised on machine intentions will look for solutions inside the machine --- better objective functions, more faithful value representations --- and will systematically underinvest in the institutional and governance structures that actually determine whether human--AI interaction remains accountable.

% ========================= Section 6
\section{Conclusion}
\label{sec:conclusion}
This paper has argued that human--LLM interaction is best understood neither through the lens of cognitive replication nor through reductive dismissal, but as asymmetric communication: a structured process in which LLMs generate contingent outputs that humans interpret, evaluate, and incorporate into ongoing language games whose normative structure remains entirely on the human side of the interaction. The framework developed in Sections~\ref{sec:philosophical-background}--\ref{sec:human-machine-communication}, through a selective composition of Wittgenstein, Luhmann, Esposito, and Brandom, provides the basis for this claim. Wittgenstein shows that meaning depends on participation in socially established practices and normative forms of assessment; Luhmann shows that communication is completed on the receiver's side; Esposito extends this insight to artificial communication; and Brandom supplies a precise account of scorekeeping, commitment, and entitlement that makes the normative asymmetry explicit. Applied to the five contested AI narratives discussed in Section~\ref{sec:ai-narratives}, this framework shows that the recurring error is the same in each case: attributing to the machine capacities that exist only within human practices of interpretation and evaluation.

The present framework also helps to interpret several recent developments in LLM engineering. For example, RLHF captures selected forms of human evaluative uptake and feeds them back into optimization as training signals~\cite{christiano_2017}. This does not make the model a scorekeeper, but formalizes aspects of receiver-side assessment while leaving normative standing on the human side. Other developments follow the same pattern. Retrieval augmentation introduces external verification into the communicative loop~\cite{lewis_2020}. Prompt engineering refines the conditions under which human users elicit coherent outputs without the model thereby following rules in any normative sense~\cite{white_2023}. Guardrails constrain the space of outputs available for uptake~\cite{dong_2024}. Agentic frameworks formalize delegated human agency mediated through computational systems rather than instituting autonomous normative participation~\cite{yao_2023}. What these advances primarily improve is not machine participation in meaning or normativity, but the conditions under which human participants evaluate outputs, verify claims, and maintain accountability. They do not show that meaning, normativity, or responsibility have shifted into the machine. On the contrary, they make visible a structural feature of the interaction: progress consists not in relocating these functions into the model, but in improving the ways human practices organize, assess, and take up its outputs.

As LLM capability increases, the human--machine pair does not converge toward autonomous AI. Instead, it deepens into more sophisticated forms of collaboration and integration. Because meaning arises through the interaction between generative artifacts and human normative practices, communication between humans and machines does not require that machines become human-like interlocutors, nor that humans reduce themselves to machine-like processors. Each side contributes according to its structural role: machines process large volumes of data at high speed and generate statistically coherent linguistic artifacts, while humans interpret, evaluate, and stabilize those artifacts within normative practices. Machines optimize throughput; humans supply discursive weight. Hence the human contribution may slow circulation, but it is what turns output into accountable communication rather than mere signal flow. However, locating the stabilization of meaning and accountability on the human side should not be mistaken for an idealization of human communicative practice. Human communities are capable not only of sustaining truth and justification but also of stabilizing error, manipulation, and strategic distortion. The human side of this asymmetry should not be understood as a single universal standpoint. Meaning and discursive authority emerge from historically situated communities whose practices vary across cultures and institutions.  The framework developed here identifies where meaning, normativity, and discursive standing are conferred, without assuming that the resulting order is epistemically reliable, morally benign, or grounded in a universal model of the human subject. The narrative that blames the AI is socially convenient because it converts human decisions into apparent machine agency. The present framework does the opposite: it restores responsibility to the human actors and institutions that design, authorize, deploy, and tolerate these systems, refusing the retrospective fiction that the machine itself became the locus of decision. Discursive standing remains a human achievement, and so does responsibility for its consequences.\footnote{\textit{Video meliora proboque, deteriora sequor} (I see and approve of the better, but I follow the worse), from the Metamorphoses Book 7, 20-1 of Ovid.}

\bibliographystyle{unsrturl}  
\bibliography{bibliography}

\end{document}